\documentclass[%
 prl,%
 amsmath,amssymb,
 reprint,%
 superscriptaddress
,floatfix
]{revtex4-1}

\usepackage{graphicx}
\usepackage{siunitx}
\usepackage[utf8]{inputenc}
\usepackage{amsmath}
\usepackage{amsfonts}
\usepackage{amssymb}
\usepackage{natbib}
\usepackage{graphicx} 
\graphicspath{{figures/}}
\makeatletter
\newcommand{\manualsublabel}[3]{(#2)\def\@currentlabel{\ref{#3}(#2)}\label{#1}}
\newcommand{\rmsub}[2]{{#1}_{\mathrm{#2}}} 
\DeclareUnicodeCharacter{FB01}{}
\DeclareUnicodeCharacter{2215}{}
\DeclareUnicodeCharacter{2212}{}
\makeatother

\begin{document}
\title{Single-photon Emission from an Acoustically-driven Lateral Light-emitting Diode}



\author{Tzu-Kan Hsiao}
\email{tzu-kan.hsiao@cantab.net (tkh28@cam.ac.uk).}
\affiliation{Department of Physics, Cavendish Laboratory, University of Cambridge, Cambridge, CB3 0HE, UK}

\author{Antonio Rubino}
\affiliation{Department of Physics, Cavendish Laboratory, University of Cambridge, Cambridge, CB3 0HE, UK}

\author{Yousun Chung}
\altaffiliation[Current address: ]{Centre of Excellence for Quantum Computation and Communication Technology, University of New South Wales, Sydney, New South Wales 2052, Australia}
\affiliation{Department of Physics, Cavendish Laboratory, University of Cambridge, Cambridge, CB3 0HE, UK}

\author{Seok-Kyun Son}
\altaffiliation[Current address: ]{National Graphene Institute, University of Manchester, Manchester, M13 9PL, UK}
\affiliation{Department of Physics, Cavendish Laboratory, University of Cambridge, Cambridge, CB3 0HE, UK}

\author{Hangtian Hou}
\affiliation{Department of Physics, Cavendish Laboratory, University of Cambridge, Cambridge, CB3 0HE, UK}

\author{Jorge Pedr\'{o}s}
\altaffiliation[Current address: ]{Instituto de Sistemas Optoelectrónicos y Microtecnología, Universidad Politécnica de Madrid,  Madrid 28040, Spain}
\affiliation{Department of Physics, Cavendish Laboratory, University of Cambridge, Cambridge, CB3 0HE, UK}

\author{Ateeq Nasir}
\affiliation{Department of Physics, Cavendish Laboratory, University of Cambridge, Cambridge, CB3 0HE, UK}
\affiliation{National Physical Laboratory, Hampton Road, Teddington, TW11 0LW, UK}

\author{Gabriel Éthier-Majcher}
\affiliation{Department of Physics, Cavendish Laboratory, University of Cambridge, Cambridge, CB3 0HE, UK}

\author{Megan J. Stanley}
\altaffiliation[Current address: ]{University of California, San Francisco, 185 Berry Street Bldg B, San Francisco CA 94158, USA}
\affiliation{Department of Physics, Cavendish Laboratory, University of Cambridge, Cambridge, CB3 0HE, UK}

\author{Richard T. Phillips}
\affiliation{Department of Physics, Cavendish Laboratory, University of Cambridge, Cambridge, CB3 0HE, UK}

\author{Thomas A. Mitchell}
\affiliation{Department of Physics, Cavendish Laboratory, University of Cambridge, Cambridge, CB3 0HE, UK}

\author{Jonathan P. Griffiths}
\affiliation{Department of Physics, Cavendish Laboratory, University of Cambridge, Cambridge, CB3 0HE, UK}

\author{Ian Farrer}
\altaffiliation[Current address: ]{Department of Electronic and Electrical Engineering, University of Sheffield, Mappin St, Sheffield, S1 3JD, UK}
\affiliation{Department of Physics, Cavendish Laboratory, University of Cambridge, Cambridge, CB3 0HE, UK}

\author{David A. Ritchie}
\affiliation{Department of Physics, Cavendish Laboratory, University of Cambridge, Cambridge, CB3 0HE, UK}

\author{Christopher J.~B. Ford}
\email{cjbf@cam.ac.uk.}
\affiliation{Department of Physics, Cavendish Laboratory, University of Cambridge, Cambridge, CB3 0HE, UK}

\begin{abstract}

Single-photon sources are essential building blocks in quantum photonic networks, where quantum-mechanical properties of photons are utilised to achieve quantum technologies such as quantum cryptography and quantum computing. Most conventional solid-state single-photon sources are based on single emitters such as self-assembled quantum dots, which are created at random locations and require spectral filtering. These issues hinder the integration of a single-photon source into a scaleable photonic quantum network for applications such as on-chip photonic quantum processors. In this work, using only regular lithography techniques on a conventional GaAs quantum well, we realise an electrically triggered single-photon source with a GHz repetition rate and without the need for spectral filtering. In this device, a single electron is carried in the potential minimum of a surface acoustic wave (SAW) and is transported to a region of holes to form an exciton. The exciton then decays and creates a single photon in a lifetime of $\sim 100$\,ps. This SAW-driven electroluminescence (EL) yields photon antibunching with $g^{(2)}(0) = 0.39 \pm 0.05$, which satisfies the common criterion for a single-photon source $g^{(2)}(0) < 0.5$. Furthermore, we estimate that if a photon detector receives a SAW-driven EL signal within one SAW period, this signal has a 79\%--90\% chance of being a single photon. 
This work shows that a single-photon source can be made by combining single-electron transport and a lateral \textit{n-i-p} junction. This approach makes it possible to create multiple synchronised single-photon sources at chosen positions with photon energy determined by quantum-well thickness. Compared with conventional quantum-dot-based single-photon sources, this device may be more suitable for an on-chip integrated photonic quantum network.
\end{abstract}

\maketitle

The development of single-photon sources is important for many quantum information technologies~\citep{Senellart2017,Aharonovich2016,Shields2007}, such as quantum cryptography~\citep{Gisin2001,Ekert1991,Lo1999}, quantum communication~\citep{Kimble2008,Pirandola2015,Horodecki2009}, quantum metrology~\citep{Giovannetti2004,Giovannetti2011}, and quantum computation~\citep{Knill2001,Kok2007}. Currently, most high-performance single-photon sources are self-assembled InGaAs-based quantum dots (QDs)~\citep{He2013,Somaschi2016,Ding2016}. However, there are several issues that may limit their integration into practical quantum photonic networks~\citep{Politi2009,Crespi2011,Harris2017,Qiang2018,Spring2013,Wang2017,Sparrow2018,Arrazola2018}. Firstly, in conventional growth of self-assembled QDs, the location and size of each QD are quite random. Therefore, one has to rely on statistics to create structures like optical cavities and gates around a quantum dot. This will be an issue for applications that require several deterministically-fabricated single-photon sources on a compact chip. Secondly, it is hard to precisely control the size of a quantum dot, which will affect the single-photon energy. Hence, it is challenging to make identical QD single-photon sources, which is essential for applications like quantum computation and a quantum repeater~\citep{Knill2001,Kaltenbaek2009}. Finally, in order to ensure that a neutral exciton is created in every optical or electrical excitation, the excitation power is usually high, which causes extra biexciton emission line. This issue can be avoided by employing resonance fluorescence, which requires a more sophisticated tuneable light source to excite each dot~\citep{Ding2016}, and needs orthogonal-excitation or cross-polarisation configurations to suppress the laser background~\citep{Ding2016,Muller2007,Ates2009}.

\begin{figure*} 
\centering    
\includegraphics[width=0.81\textwidth]{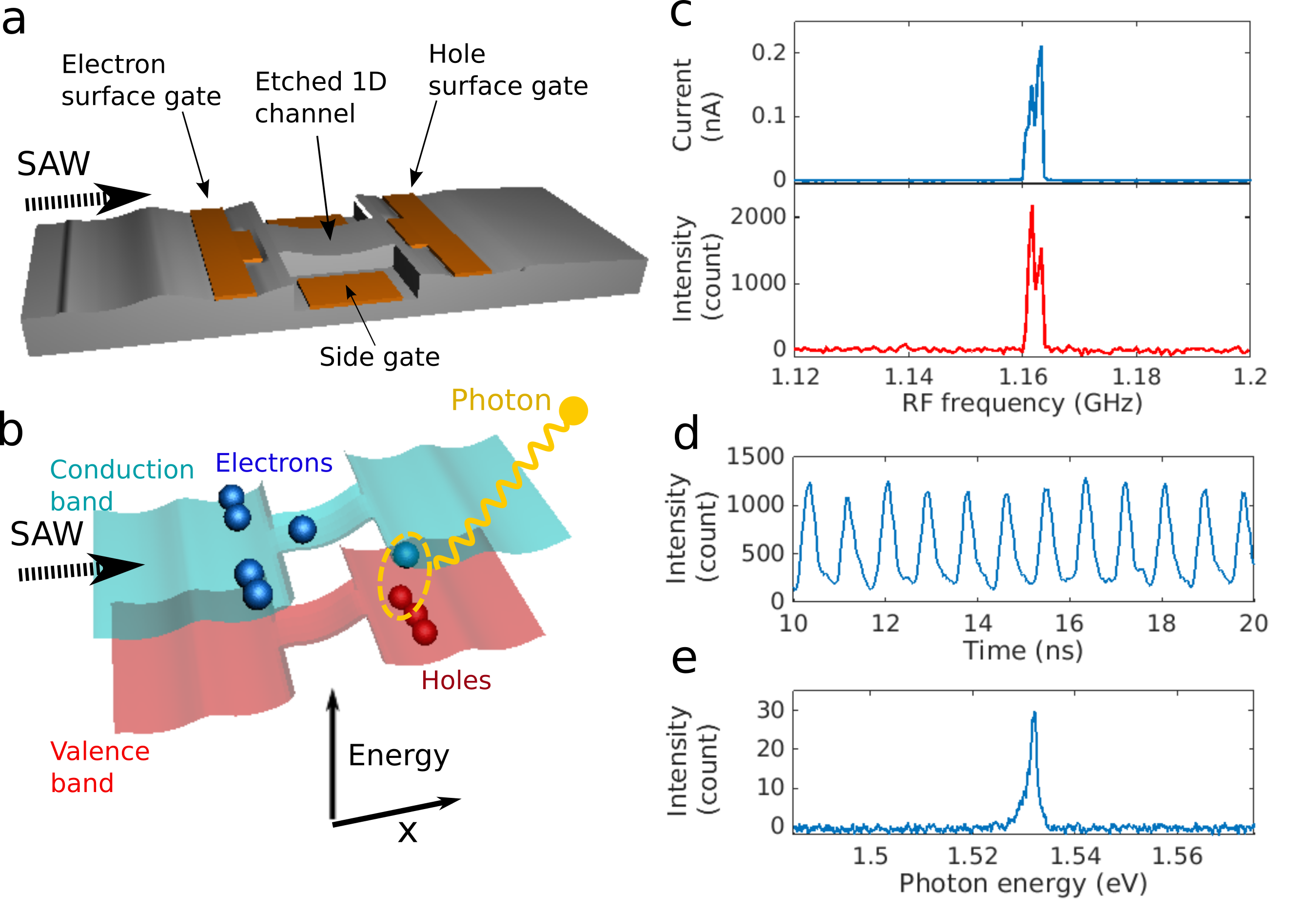}
\caption{SAW-driven lateral \textit{n-i-p} junction, and its electrical and optical properties. (a) Schematic of the device. Electron and hole surface gates induce electrons and holes in a GaAs quantum well, forming a lateral \textit{n-i-p} junction along an etched 1D channel. A SAW is generated by applying an RF signal to a transducer (placed 1\,mm from the \textit{n-i-p} junction). (b) Schematic diagram showing the band structure of the \textit{n-i-p} junction modulated by the SAW potential, for an applied forward bias less than the bandgap. A single electron is carried in each SAW minimum, producing a single photon when it recombines with a hole. (c) S-D current (top) and EL intensity (bottom) as a function of applied RF frequency at an RF power of 9\,dBm. They both show a significant enhancement around 1.163\,GHz, which is the resonant SAW frequency of the IDT. (d) SAW-driven EL intensity as a function of time. The 860\,ps periodic feature corresponds to the applied SAW frequency of 1.163\,GHz. (e) Energy spectrum of the SAW-driven EL shown in (c). The spectrum shows a peak at 1.531\,eV (FWHM $\sim 1$\,meV), which matches the exciton energy in the quantum well (see Supplementary Information S1).}
\label{fig_device}
\end{figure*}

On the other hand, many prototypical quantum processors, for tasks such as boson sampling, Shor's algorithm, arbitrary two-qubit gates, and quantum molecular-dynamics simulation, have recently been realised using on-chip photonic quantum networks~\citep{Politi2009,Crespi2011,Harris2017,Qiang2018,Spring2013,Wang2017,Sparrow2018}. In these works, multiple heralded single-photon states are prepared via spontaneous parametric down-conversion or spontaneous four-wave mixing, which both have low efficiency issue due to their Poissonian photon-number distribution~\citep{Migdall2002,Collins2013}.
Therefore, if a single-photon source, which exhibits sub-Poissonian photon statistics, can be fabricated just by conventional lithography techniques, it will be possible to make multiple sources at predefined locations on a photonic chip. Moreover, if these sources can be electrically triggered at a GHz-repetition rate, an ideal and compact photonic quantum processor may thus be achieved.

To try to overcome the issues of the randomness of self-assembled quantum dots and to create integrated on-chip photonic quantum networks, a single-photon source can in principle be made in a novel approach combining a single-electron-transport technique with a lateral \textit{n-i-p} junction, a semiconductor interface between adjacent electron (`\textit{n}') and hole (`\textit{p}') regions, formed in a quantum well, with an intrinsic region (`\textit{i}') in between (a lateral light-emitting diode). In recent years, control of a propagating single-electron quantum state has been achieved using an electron pump \citep{Giblin2012}, a leviton \citep{Dubois2013}, and a surface acoustic wave (SAW)~\citep{Shilton1996,PhysRevB.56.15180,Ford2017,Bauerle2018}. If a single electron can be transported from an electron region into a region of holes (an \textit{n-i-p} junction), it can form an exciton and then decay into a single photon. In this approach, the emission of single photons is caused by the single-electron transport, rather than the stochastic capture of electrons in a static quantum dot. Therefore, it does not suffer from the spectral and temporal randomness of self-assembled quantum-dots, and is possible to be made at any arbitrary location using conventional lithography techniques. Moreover, because only one exciton can be formed at a time using the single-electron-transport technique, a biexciton will not form in this approach, eliminating the need for extra spectral filtering. Furthermore, since the exciton energy is determined by the thickness of the quantum well, the single-photon energy can be easily predicted, which makes the optical-cavity integration more reliable.

Based on the approach mentioned above, a scheme for making a single-photon source using a SAW has been proposed~\citep{Foden2000}. This makes use of the fact that, in a piezoelectric material such as GaAs, a SAW consists of both a strain and a potential modulation. In a narrow channel, electrons are confined in moving quantum dots formed by the SAW potential and the lateral channel potential. The number $n$ of electrons in each SAW potential minimum is well defined if the Coulomb charging energy is large enough. The SAW (of frequency $f_{\rm SAW}$) can therefore drive a quantised current $nef_{\rm SAW}$ along the channel ($e$ is the electronic charge) \cite{Shilton1996,Ford2017}. To generate light, single electrons must be carried in SAW potential minima across a lateral \textit{n-i-p} junction to create single photons by recombining with holes (see Supplementary Video for a simple animation). Over more than a decade, various attempts have been made to implement this scheme~\citep{Hosey2004,Cecchini2005,Smith2006,Gell2006,DeSimoni2009,Dai2014}. However, to the authors' knowledge, no evidence for photon antibunching or single-photon emission has been observed. Note that a few works based on SAW-injected or SAW-regulated excitons in a QD have shown single-photon emission~\citep{Iikawa2009,Weib2016,Villa2017}. However, they still rely on the presence of a QD and thus suffer from the randomness issue. Here we successfully demonstrate that an electrically triggered and GHz-repetition-rate single-photon source based on this scheme can be fabricated in a deterministic conventional lithography process, and, without any spectral filtering, our device shows clear photon antibunching with the second-order correlation function $g^{(2)}(0)=0.39 \pm 0.05$, which is lower than the common criterion for a single-photon source, $g^{(2)}(0)<0.5$.

In this work, the lateral \textit{n-i-p} junction is made in a conventional undoped GaAs quantum well using standard lithography techniques (see Methods). Electrons and holes are induced in the regions under the electron and hole surface gates, which are separated by an intrinsic region (Fig.\ \ref{fig_device}(a)). A source-drain (S-D) bias less than the GaAs bandgap is applied across the \textit{n-i-p} junction to create a finite potential difference between the electron and hole regions. A SAW is generated by applying a radio-frequency (RF) signal to an interdigitated transducer (IDT) at its resonant frequency $f_{\rm SAW}=1.163$\,GHz. Electrons are trapped in SAW potential minima and pushed towards the hole region (Fig.\ \ref{fig_device}(b)). To achieve SAW-driven single-electron transport, lateral confinement is provided by etching the region connecting the electron and hole regions into a 1D intrinsic channel. In addition, a pair of side gates is placed on either side of the channel to adjust the electrostatic potential in the intrinsic region. The physical length of the channel is made to be similar to the SAW wavelength of 2.5\,$\mu$m. In this case, 
any current flow will be caused by the SAW carrying electrons up the potential slope linking the conduction band in the regions of electrons and holes, not by the SAW reducing the height of the potential barrier in the intrinsic region at a certain part of its cycle. All measurements were carried out at 1.5\,K.

In order to test the effect of a SAW on the induced lateral \textit{n-i-p} junction, a S-D bias $V_{\rm SD}$ less than 1.45\,V is applied to the junction. This is at least 90\,mV below the voltage required to align the conduction band in the \textit{n} and \textit{p} regions so that current can flow at cryogenic temperature if any intermediate barrier is overcome. In this case, due to the conduction-band offset between the \textit{n} and \textit{p} regions, electrons cannot reach the \textit{p} region to recombine with holes unless a SAW carries them there. Therefore, a S-D current and electroluminescence (EL) signal will only appear when an RF signal is applied to the IDT at its resonant frequency $f_{\rm SAW}$. Examples of the SAW-driven current and EL are shown in Fig.\ \ref{fig_device}(c). The S-D current (Fig.\ \ref{fig_device}(c) top panel) is greatly enhanced around $f_{\rm SAW} \sim 1.163$\,GHz with an RF power of 9\,dBm (quality factor $\frac{f_{\rm SAW}}{\Delta f} \sim 390$). This SAW-driven current is close to $1\,ef_{\rm SAW} = 0.186$\,nA. It means that the number of electrons carried in each SAW minimum is roughly one on average, a single-electron regime which will, in principle, generate single photons. These electrons driven by the SAW arrive at the hole region and recombine, causing a SAW-driven EL signal, as seen in Fig.\ \ref{fig_device}(c) (bottom panel). The EL signal is emitted from the \textit{p} region as electrons recombine with holes there. The internal quantum efficiency $\eta$, defined as the ratio of the number of photons actually collected to the number of photons that can theoretically be collected by the optics, is about 2.5\%. This low $\eta$ may be caused by trapping and non-radiative recombination in surface states around the etched edges~\citep{Demichel2010,Shiozaki2005}, or due to electrons being carried away without recombining near the junction. Time-resolved measurements of the SAW-driven EL, shown in Fig.\ \ref{fig_device}(d), exhibit periodic peaks with a period of 860\,ps. Hence, it is evident that electrons are injected into the hole region by the SAW, leading to photon emission with the period of the SAW.

The spectrum of the SAW-driven EL is shown in Fig.\ \ref{fig_device}(e). The spectral peak corresponds to the neutral-exciton transition from the conduction band to the first heavy-hole subband in the quantum well (see Supplementary Information S1)~\citep{Hegarty1985}. The full width at half maximum (FWHM) of the peak is about 1\,meV, which can be attributed to acoustic-phonon scattering ($\Delta E \sim 0.2$\,meV at 1.5\,K) and interface roughness (atomic monolayer fluctuations in the quantum-well thickness give $\Delta E \sim 0.5$\,meV)~\citep{Srinivas1992,Singh2003}. The lower-energy tail of the peak may be due to localised exciton states or a Stark shift in the hole region. Unlike conventional single-photon sources based on self-assembled quantum dots, which usually have an extra peak in the spectrum due to biexciton states, this device shows only one peak (neutral exciton) without any spectral filtering or optical cavity. Hence, it should be possible to obtain single-photon emission from this SAW-driven lateral \textit{n-i-p} junction even though there is no spectral filtering or optical cavity.

\begin{figure}[t!] 
\centering    
\includegraphics[width=\columnwidth]{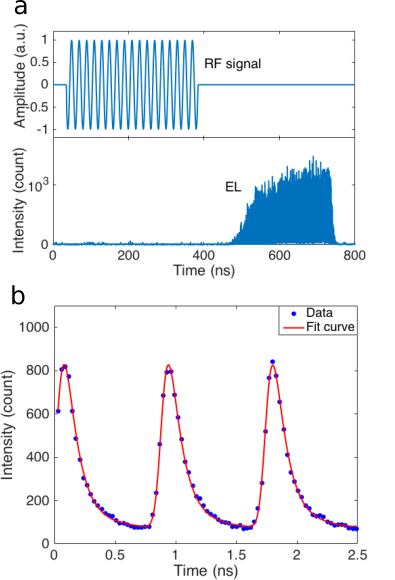}
\caption{Dynamics of SAW-driven electron transport. (a) A 350\,ns-long pulsed RF signal (top, shown at a low frequency for clarity) is applied to the IDT to create a pulsed SAW. The SAW-driven EL signal (bottom) is delayed by roughly 400\,ns owing to propagation of the SAW from the IDT to the \textit{n-i-p} junction. (b) Averaged SAW-driven EL and the best fit using $H(t)$ (see Supplementary Information S3).}
\label{fig_time_resolved}
\end{figure}

The dynamics of the SAW-driven electron transport can be studied using a time-resolved EL measurement technique. A 350\,ns-long pulsed RF signal is applied to the IDT to generate a pulsed SAW (Fig.\ \ref{fig_time_resolved}(a) top). The SAW-driven current is close to the single-electron regime. Because the SAW velocity on GaAs is $\sim 2800$\,m/s and the distance from the IDT to the \textit{n-i-p} junction is $\sim 1.1$\,mm, it will take about 400\,ns for the SAW to arrive at the junction, and for its amplitude to build up so that it transports electrons which then recombine with holes. Therefore, compared with the RF signal, SAW-driven EL will be delayed by about 400\,ns, as can be seen in Fig\ \ref{fig_time_resolved}(a) (bottom). This confirms that the EL signal is indeed caused by the SAW, rather than by electromagnetic (EM) crosstalk generated by the RF signal, which should have an effect without any noticeable delay since the speed of light is five orders of magnitude faster than the SAW.

In order to understand more detailed dynamics, data points three SAW periods apart are averaged across a large part of the region where the EL signal is observed, in Fig.\ \ref{fig_time_resolved}(a) (bottom), to give three periods that are the combination of every third period of the data. The resulting data is shown in Fig.\ \ref{fig_time_resolved}(b). The shape of an individual peak can be understood from the injection of electrons by the SAW. When an electron is pumped across the \textit{n-i-p} junction by the SAW, the probability of electron-hole recombination suddenly steps up and causes a rapid enhancement of the EL signal. The signal then decays exponentially as the probability that the electron has already recombined rises. These peaks in Fig.\ \ref{fig_time_resolved}(b) are broadened by the temporal uncertainty (jitter) of the single-photon avalanche photodiode (SPAD) and of the SAW-driven electron transport itself, originating from a slight uncertainty about the position of an electron in a SAW minimum. Note that each peak in Fig.\ \ref{fig_time_resolved}(b) does not decay to zero by the time the next peak appears. One reason for this non-zero background level, $BG_{\rm EL}$, is that the lifetime of the dominant exciton state is not short enough. Hence, there is significant overlap between two consecutive peaks. Other reasons may be an effective background due to after-pulsing of the SPAD~\citep{Ziarkash2018} or to slowly-decaying secondary-exciton states (lifetime $\sim 0.2$--$1.5$\,ns)~\citep{Feldmann1987,Angeloni1989}. These slowly-decaying exciton states may be the localised excitons seen in the small lower-energy tail in Fig.\ \ref{fig_device}(e). In Fig.\ \ref{fig_time_resolved}(b), dynamical parameters, including carrier lifetime $\tau$ and jitter $w$, can be quantified by fitting the data to a theoretical function $H(t)$ describing the SAW-driven EL (see Supplementary Information S3). The best fit, plotted along with the data in Fig.\ \ref{fig_time_resolved}(b), gives $\tau=94$\,ps, $w=33$\,ps, and $\rmsub{BG}{EL} = 7$\% of the peak height. The 94\,ps carrier lifetime of SAW-driven electrons is short compared with the 860\,ps SAW period, so photons driven by consecutive SAW minima will not overlap significantly in the time domain, which is desirable for a single-photon source.

\begin{figure}[t!] 
\centering    
\includegraphics[width=\columnwidth]{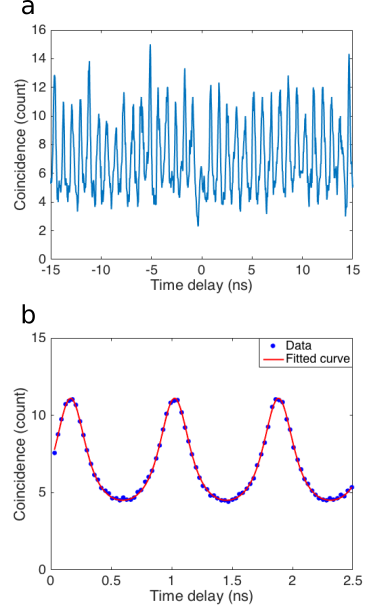}
\caption{Photon antibunching in SAW-driven EL. (a) Autocorrelation histogram of the SAW-driven EL. The coincidence at $\Delta t=0$ is suppressed to about a half of the average peak value, indicating photon antibunching, i.e.\ that there is a reduced probability of two photons arriving simultaneously. (b) Averaged autocorrelation histogram and the best fit using $G(\Delta t)$ (see Supplementary Information S4).}
\label{fig_autocorrelation}
\end{figure}

In this device, quantised SAW-driven current cannot be observed, meaning that there is some variation in the number of electrons in each SAW minimum. However, the probability distribution of electron occupation numbers can still be affected by the discrete nature of SAW-driven charge transport, causing a reduced variance in electron number. The probability distribution should thus become a sub-Poissonian distribution, which will lead to photon antibunching after recombination. Photon antibunching in the SAW-driven EL is tested by measuring an autocorrelation histogram using a Hanbury Brown and Twiss (HBT) setup (see Methods). A continuous SAW is used to drive the \textit{n-i-p} junction in the single-electron regime (with an average number of electrons in a SAW minimum of 0.89) stablised by a feedback control loop (see Supplementary Information S2). The autocorrelation histogram in Fig.\ \ref{fig_autocorrelation}(a) shows periodic peaks with the 860\,ps SAW period, indicating that coincidences in the histogram are indeed caused by the periodic SAW-driven photon emission. The peak at $\Delta t=0$ is suppressed to 65\% of the average peak value. The suppression at $\Delta t=0$ is clear evidence of photon antibunching in the SAW-driven EL (a reduced probability of two photons arriving at the same moment), demonstrating that the SAW-driven \textit{n-i-p} junction can operate as a quantum photon emitter.

Although photon antibunching is observed in Fig.\ \ref{fig_autocorrelation}(a), the second-order correlation function $g^{(2)}(\Delta t)$, which confirms the presence of single-photon emission if $g^{(2)}(0) < 0.5$, cannot be simply obtained from the peak heights. This is because coincidence at a peak can have a contribution from the two neighbouring peaks if they have significant overlap, and also because an effective background ($BG_{\rm EL}$) in EL can give rise to a background ($BG_{g2}$) in the autocorrelation histogram. Therefore, the actual shape of individual peaks and the background $BG_{g2}$ have to be considered in order to extract the real $g^{(2)}(\Delta t)$. The peak shape and the background can be estimated by fitting the autocorrelation histogram to a theoretical function $G(\Delta t)$ describing the autocorrelation of SAW-driven EL (see Supplementary Information S4). To have a better fit, points from every third peak in the histogram are averaged together. The averaged histogram and the best fit are plotted in Fig.\ \ref{fig_autocorrelation}(b). The fit shows that the autocorrelation histogram is caused by a SAW-driven EL signal with $\tau=99$\,ps, $w=33$\,ps and $BG_{\rm EL}=8$\% of the peak height. These parameters are consistent with those obtained in fitting the time-resolved data (Fig.\ \ref{fig_time_resolved}(b)). With these parameters, the actual shape of individual peaks and the background $BG_{g2}$ are known. Hence, the real $g^{(2)}(\Delta t)$ can now be extracted from the autocorrelation histogram.

\begin{figure}[t!] 
\centering    
\includegraphics[width=\columnwidth]{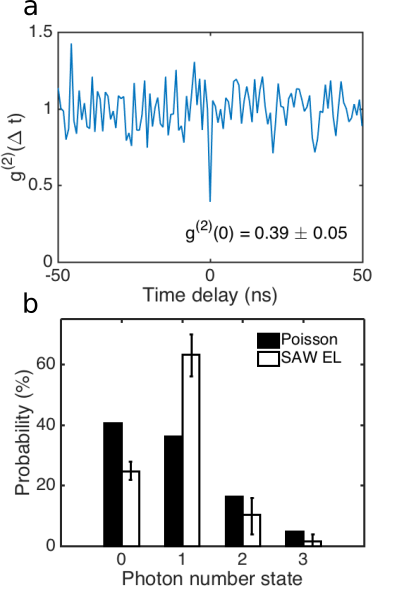}
\caption{SAW-driven single-photon emission. (a) The second-order correlation function $g^{(2)}(\Delta t)$, which is obtained after fitting the autocorrelation histogram to $G'(\Delta t)$. $g^{(2)}(0)=0.39 \pm 0.05 < 0.5$ shows that the SAW-driven lateral \textit{n-i-p} junction can produce single-photon emission. (b) Estimated probability distribution of photon-number states $\lvert n \rangle$ in the SAW-driven EL, compared with the probability distribution in a Poissonian light source with $N_{\rm avg}=0.89$.}
\label{fig_g2}
\end{figure}

$g^{(2)}(\Delta t)$ of the SAW-driven EL is obtained by finding the real contribution from each peak in the autocorrelation histogram (see Supplementary Information S5). The result is shown in Fig.\ \ref{fig_g2}(a). The suppressed photon-antibunching peak at $\Delta t=0$ gives $g^{(2)}(0)=0.39 \pm 0.05$, showing that the SAW-driven \textit{n-i-p} junction can indeed produce single-photon emission. In addition, since the average number of electrons, $N_{\rm avg}$, in a SAW minimum is 0.89, the probability distribution of photon-number states can be estimated. The wave function of electrons in a SAW minimum can be expressed in the Fock basis
\begin{equation}
\lvert \psi \rangle =\sqrt{P_{0}}\lvert 0 \rangle+\sqrt{P_{1}}\lvert 1 \rangle+\sqrt{P_{2}}\lvert 2 \rangle+ \sqrt{P_{3}}\lvert 3 \rangle+\cdots
\end{equation}
where $\lvert n \rangle$ and $P_n$ denote the electron-number states and their respective probabilities. $N_{\rm avg}$ is thus a function of the probability distribution \{$P_n$\}. When $n$ electrons (electron-number state $\lvert n \rangle$) arrive at the hole region, each of these electrons may recombine with a hole and produce a photon according to the internal quantum efficiency $\eta$. Hence, up to $n$ photons will be produced from $\lvert n \rangle$. These photons will then cause coincidences in an autocorrelation histogram. As a result, $g^{(2)}(0)$ is also a function of the probability distribution \{$P_n$\}. Assuming that $\lvert \psi \rangle$ has no projection to $\lvert m \rangle$ with $m \geq 4$ (meaning that a SAW minimum can carry only up to three electrons) and given that $N_{\rm avg}=0.89$ and $g^{(2)}(0)=0.39$, a probability distribution \{$P_0, P_1, P_2, P_3$\} of electron-number states (and photon-number states) in the SAW-driven \textit{n-i-p} junction can be estimated to be \{$0.25 \pm 0.03, 0.63 \pm 0.07, 0.10 \pm 0.06, 0.02 \pm 0.02$\} (see Supplementary Information, section S6). This probability distribution is shown in Fig.\ \ref{fig_g2}(b), along with the distribution expected for a classical Poissonian light source (with the same average number $N_{\rm avg}=0.89$) for comparison. It can be seen that, in the SAW-driven \textit{n-i-p} junction, the single-photon probability is greatly enhanced compared with the classical case. Based on this estimated probability distribution, when a detector receives a light signal from this SAW-driven \textit{n-i-p} junction, the signal has a probability of $P_1/(P_1 + P_2 + P_3) = 0.79$--$0.9$ to actually be a single photon.

Although photon antibunching and single-photon emission have been experimentally demonstrated in this SAW-driven single-photon source, many improvements need to be done in future devices. In order to understand how to enhance $P_1$, we build a simplified SAW-transport model to calculate the probability distribution of SAW-driven electrons. The result indicates that a more well-defined single-electron state (and thus higher $P_1$) may be achieved with a stronger confinement in the 1D channel and the SAW potential (see Supplementary Information S7). This may be done by using a narrower channel, and depositing ZnO thin film to enhance the SAW potential~\citep{Pedros2011}. As for the low efficiency $\eta$, this can be improved by surface passivation to reduce non-radiative surface states~\citep{Shiozaki2005}, and by better capturing of the SAW-driven electrons and building an optical cavity (see Supplementary Information S8). On the other hand, to be useful for a quantum repeater or for quantum computing, good photon indistinguishability is important. The SAW-driven EL shows a peak with FWHM $\sim 1$\,meV, which is two orders of magnitude larger than the spectral FWHM from a self-assembled QD, but comparable to the FWHM of single photons in photonic quantum networks~\citep{Tanida2012,Crespi2013,Spring2013,Wang2017,Sparrow2018}. It has been shown experimentally that single photons with a FWHM $\sim 1$\,meV can show good photon indistinguishability within a coherence time of 0.5-1\,ps (coherent length $\sim 300$\,$\mu$m) and thus can be used to perform boson sampling and quantum simulation on integrated photonic chips. Hence, multiple SAW-driven single-photon sources may be integrated into a GaAs-based photonic waveguide chip~\citep{FattahPoor2013,Prtljaga2014,Jons2015}, to achieve electrically-driven photonic quantum computation. Moreover, the photon indistinguishability in our device may be improved by using a high-quality quantum-well wafer with improved interface roughness (see Supplementary Information S8).

In this work, we have shown that a single-photon source can be made from a SAW-driven lateral \textit{n-i-p} junction fabricated in a deterministic lithographic process using gates, etching and an interdigitated transducer. Hence, multiple single-photon sources at the same energy (limited by interface roughness) could be made in any arbitrary arrangement, which is beneficial for realising on-chip integrated photonic networks. On the other hand, since growing quantum wells with highly precise thickness is possible using molecular-beam epitaxy (MBE), and the single-photon energy in this device depends on the quantum-well thickness, making identical single-photon sources using this scheme is arguably easier than using self-assembled quantum dots. This is useful for applications that require photon entanglement. Finally, this device structure may also be fabricated using emerging 2D materials like WSe\textsubscript{2} and MoS\textsubscript{2} to realise more advanced quantum photon emitters~\citep{Mak2012,Zhang2014a}. In these 2D materials, photon polarisations are coupled to the K and K' valleys in their band structures owing to the breaking of inversion symmetry. Because the junction direction and the SAW direction can both be easily reversed in such a device structure, a chiral-single-photon source with electrically-controlled polarisation may be achieved using a similar device design, which will be useful in various quantum information technologies.

\section{Methods}

\subsection{Device Fabrication}

The SAW-driven lateral \textit{n-i-p} junction was made in a 15\,nm undoped GaAs quantum well. The quantum-well layer structure consists of (from the top) a 10\,nm GaAs capping layer, a 100\,nm Al\textsubscript{0.33}Ga\textsubscript{0.67}As top barrier, a 15\,nm GaAs quantum well, a 285\,nm Al\textsubscript{0.33}Ga\textsubscript{0.67}As back barrier, and finally a 1\,$\mu$m GaAs buffer layer. \textit{n}-type and \textit{p}-type ohmic contacts were in direct contact with the quantum well. AuGeNi (for \textit{n}-type contacts) and AuBe (for \textit{p}-type contacts) were evaporated in regions recessed to the quantum well, and annealed at \SI{470}{\degreeCelsius} and \SI{520}{\degreeCelsius} respectively. Bridging gates for inducing electrons and holes from the ohmic contacts were fabricated by evaporating Ti/Au on a 100\,nm Al\textsubscript{2}O\textsubscript{3} insulator, which is on top of the ohmic contacts. Surface gates for extending charge carriers and forming a lateral \textit{n-i-p} junction were fabricated on the wafer surface using electron-beam lithography. 1.2\,$\mu$m-wide 1D channel between the electron and hole regions was made by removing quantum well next to the 1D channel using electron-beam-defined wet etching. The interdigitated transducers with a period of 2.5\,$\mu$m were made using electron-beam lithography. The completed device was then mounted on a custom-made sample holder for measurement at 1.5\,K.

\subsection{Optical Setup}

A EL signal emitted from a 2\,$\mu$m\textsuperscript{2} area is collected by a home-made confocal fibre-coupled lens assembly, the position of which is controlled by a three-axis piezoelectric stage relative to the sample. The EL signal is then sent through a single-mode fibre to optical components outside the cryostat. The EL signal was spectrally filtered by a 750mm Czerny-Turner spectrometer before being detected by a chilled EMCCD camera.

\subsection{Time-resolved EL Measurement}

A time-resolved EL setup consists of a SPAD, an arbitrary waveform generator that produces timing pulses, a time-to-digital converter, and a RF source that is synchronised to the timing pulses. The SPAD can be triggered by the detection of a single photon and will then output a signal pulse with a 40\,ps jitter. The pulse generator also produces timing pulses with a 10\,ps jitter. The signal pulses and the timing pulses are connected to the time-to-digital converter, which measures the time difference between a timing pulse and a signal pulse. The RF source is synchronised with the pulse generator by using a 10\,MHz sync signal. The RF signal is used to generate a SAW signal synchronised to the timing pulses. A time-correlated histogram of a SAW-driven EL signal can thus be obtained by recording the time difference between timing pulses and SPAD signal pulses.

\subsection{Hanbury Brown-Twiss Setup}

An HBT setup consists of a 50:50 fibre-coupled beam-splitter and two SPADs. The beam-splitter splits the stream of photons in a SAW-driven EL signal into two beams. Each beam is sent to a SPAD. These two SPADs produce signal pulses when they are triggered by the incoming photons. In a start-stop autocorrelation method, a signal pulse from SPAD 1 (start) will cause the time-to-digital converter to begin time counting until the counter receives a signal pulse from SPAD 2 (stop). The counter then records one coincidence at the time delay between these two signal pulses. The two beams of photons will give rise to the autocorrelation histogram, which is the number of coincidences as a function of time delay.

\bibliography{arxiv_main_tex.bbl}

{\bf Acknowledgements} This work was supported by the European Union Horizon 2020 research and innovation program under Marie Sk\l{}odowska-Curie Grant Agreement No. 642688 (SAWtrain), and the UK EPSRC [Grant Nos. EP/J003417/1 and EP/H017720/1]. A.N. was supported by the UK Department for Business, Innovation and Skills. TKH was supported by the Cambridge Overseas Trust. JP acknowledges the support from the Spanish MICINN (Programa Nacional de Movilidad de Recursos Humanos I+D+i) and from the 7th European Community Framework Programme (Marie Curie Intra-European Fellowship, Project no. 235515) during his time at the University of Cambridge, as well as from the Spanish MINECO (grant RyC-2015-18968) now at Universidad Politécnica de Madrid. We thank Mete Atat\"{u}re for the loan of the SPADs and the time-to-digital converter and Charles Smith and Joanna Waldie for the loan of the arbitrary waveform generator. We also thank Peter Spencer for assistance with the optical setup.

{\bf Author contributions} Project planning: CJBF, TKH and JP; MBE growth: IF and DAR; E-beam lithography: JPG and TAM; sample design and fabrication: TKH, YC, AR, SKS and AN; electrostatic calculation: HH; scanning cryogenic microscope: JP, YC, RTP, TKH, SKS, HH and CJBF; Spectrometer, HBT and time-resolved setup: AN, GÉM, MJS and RTP; transport and optical measurements: TKH, AN and AR; analysis of results and theoretical interpretation: TKH and CJBF.

{\bf Data availability} Data associated with this work are available at the University of Cambridge data repository (https://doi.org/???; see ref. ??).

\textbf{Competing Interests} The authors declare that they have no competing financial interests.

\textbf{Correspondence} Correspondence and requests for materials
should be addressed to C.\ J.\ B.\ Ford (email: cjbf@cam.ac.uk) or Tzu-Kan Hsiao (email: tzu-kan.hsiao@cantab.net or tkh28@cam.ac.uk).

\end{document}


\title{Single-photon Emission from an Acoustically-driven Lateral Light-emitting Diode}



\author{Tzu-Kan Hsiao}
\email{tzu-kan.hsiao@cantab.net (tkh28@cam.ac.uk).}
\affiliation{Department of Physics, Cavendish Laboratory, University of Cambridge, Cambridge, CB3 0HE, UK}

\author{Antonio Rubino}
\affiliation{Department of Physics, Cavendish Laboratory, University of Cambridge, Cambridge, CB3 0HE, UK}

\author{Yousun Chung}
\altaffiliation[Current address: ]{Centre of Excellence for Quantum Computation and Communication Technology, University of New South Wales, Sydney, New South Wales 2052, Australia}
\affiliation{Department of Physics, Cavendish Laboratory, University of Cambridge, Cambridge, CB3 0HE, UK}

\author{Seok-Kyun Son}
\altaffiliation[Current address: ]{National Graphene Institute, University of Manchester, Manchester, M13 9PL, UK}
\affiliation{Department of Physics, Cavendish Laboratory, University of Cambridge, Cambridge, CB3 0HE, UK}

\author{Hangtian Hou}
\affiliation{Department of Physics, Cavendish Laboratory, University of Cambridge, Cambridge, CB3 0HE, UK}

\author{Jorge Pedr\'{o}s}
\altaffiliation[Current address: ]{Instituto de Sistemas Optoelectrónicos y Microtecnología, Universidad Politécnica de Madrid,  Madrid 28040, Spain}
\affiliation{Department of Physics, Cavendish Laboratory, University of Cambridge, Cambridge, CB3 0HE, UK}

\author{Ateeq Nasir}
\affiliation{Department of Physics, Cavendish Laboratory, University of Cambridge, Cambridge, CB3 0HE, UK}
\affiliation{National Physical Laboratory, Hampton Road, Teddington, TW11 0LW, UK}

\author{Gabriel Éthier-Majcher}
\affiliation{Department of Physics, Cavendish Laboratory, University of Cambridge, Cambridge, CB3 0HE, UK}

\author{Megan J. Stanley}
\altaffiliation[Current address: ]{University of California, San Francisco, 185 Berry Street Bldg B, San Francisco CA 94158, USA}
\affiliation{Department of Physics, Cavendish Laboratory, University of Cambridge, Cambridge, CB3 0HE, UK}

\author{Richard T. Phillips}
\affiliation{Department of Physics, Cavendish Laboratory, University of Cambridge, Cambridge, CB3 0HE, UK}

\author{Thomas A. Mitchell}
\affiliation{Department of Physics, Cavendish Laboratory, University of Cambridge, Cambridge, CB3 0HE, UK}

\author{Jonathan P. Griffiths}
\affiliation{Department of Physics, Cavendish Laboratory, University of Cambridge, Cambridge, CB3 0HE, UK}

\author{Ian Farrer}
\altaffiliation[Current address: ]{Department of Electronic and Electrical Engineering, University of Sheffield, Mappin St, Sheffield, S1 3JD, UK}
\affiliation{Department of Physics, Cavendish Laboratory, University of Cambridge, Cambridge, CB3 0HE, UK}

\author{David A. Ritchie}
\affiliation{Department of Physics, Cavendish Laboratory, University of Cambridge, Cambridge, CB3 0HE, UK}

\author{Christopher J.~B. Ford}
\email{cjbf@cam.ac.uk.}
\affiliation{Department of Physics, Cavendish Laboratory, University of Cambridge, Cambridge, CB3 0HE, UK}

\maketitle

\onecolumngrid


\section{Supplementary Information}

\subsection{S1. Exciton Energy}

\begin{figure}[b]
\centering    
	\includegraphics[width=0.7\textwidth]{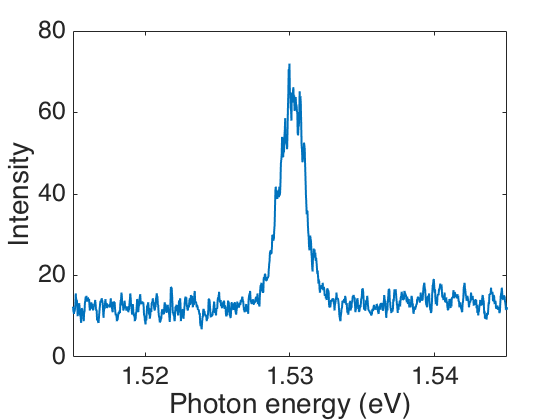}
\caption{Photoluminescence (PL) spectrum of the 15\,nm GaAs quantum well at 1.5\,K.}
\label{fig_PL}
\end{figure}

Energy of the neutral exciton in this 15\,nm-quantum-well structure can be characterised by photoluminescence (PL). Neutral excitons are created in the quantum well by optical excitation using a 645\,nm laser diode. Fig.\ \ref{fig_PL} shows the PL spectrum of the 15\,nm GaAs quantum well, taken in the \textit{n-i-p}-junction region at 1.5\,K. The main 1.530\,eV peak (FWHM $\sim 2$\,meV) is due to the excitonic transition of electrons from the conduction band to the first heavy-hole subband. The peak in the spectrum was also observed in EL signals from multiple lateral \textit{n-i-p} junctions based on the same 15\,nm quantum well, indicating that the EL signals are dominated by the neutral excitons. 

\subsection{S2. Stabilising SAW-driven S-D Current}

Ideally, the device should be operated precisely in the single-electron regime (S-D current $=1ef$). However, even with the S-D bias and all gate voltages kept fixed, the SAW-driven S-D current can drift between $5ef$ (multiple-electron regime) and $0ef$ (vacuum state). This instability may be related to a charging effect near the \textit{n-i-p} junction so that the potential drifts and affects the SAW-driven charge transport. In order to deal with this instability, a PID control loop was used so that constant adjustment could be made to S-D bias to keep the SAW-driven current around $0.9ef$. Note that the SAW-driven current still occasionally drifts above $1ef$ or below $0.8ef$, so the coincidences that occurred at these moments are removed from the autocorrelation histogram in order to show the $g^{(2)}(\Delta t)$ when the SAW-driven charge transport is in the single-electron regime.

\subsection{S3. Fitting for Time-resolved EL Data}

\begin{figure}
\centering    
	\includegraphics[width=0.7\textwidth]{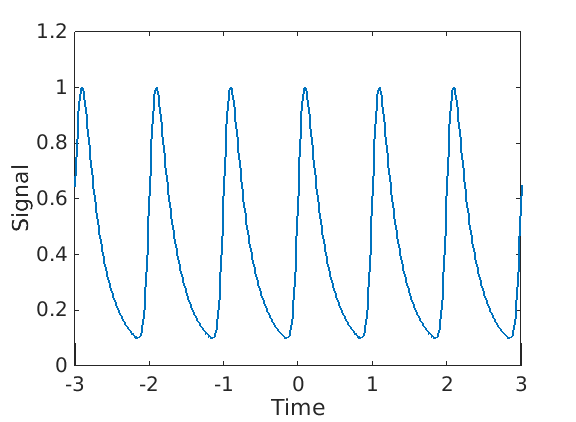}
\caption{Example of function $H(t)$ with $t_0 = 1$, $\rmsub{A}{EL} = 1$, $\tau = 0.2$, $w = 0.05$, $\rmsub{t}{SAW} = 1$, and $\rmsub{BG}{EL} = 0$}.
\label{fit_time_resolved_func}
\end{figure}

A theoretical function describing the SAW-driven EL can be established to fit the time-resolved measurement results. Because the SAW-driven charge transport gives rise to a sudden increase in the density of electrons in the region of holes, and the density of electrons then decays due to the electron-hole recombination, the basic function for each peak in a time-correlated histogram will be
\begin{align}
f(t,t_0,\rmsub{A}{EL},\tau) & = 0 & \text{when} \quad t & < t_0 \,;\notag \\
					& = \rmsub{A}{EL} \cdot \exp((t-t_0)/\tau) & \text{when} \quad t & \geq t_0 \,,
\end{align}
where $t$ is time, $t_0$ is the arrival time of the SAW potential minimum leading to the peak, $\rmsub{A}{EL}$ is the amplitude (peak value), and $\tau$ is the carrier lifetime of SAW-driven electrons. Since the EL is driven by a series of SAW potential minima, the overall function can be written as
\begin{equation}
F(t,t_0,\rmsub{A}{EL},\tau,\rmsub{t}{SAW}) = \sum_{i = 0,1,2,3\ldots} f(t,{t_0}+i \cdot \rmsub{t}{SAW},\rmsub{A}{EL},\tau) \,,
\label{eq_F_function}
\end{equation}
where $\rmsub{t}{SAW}$ is the period of the SAW. In order to fit experimental data, temporal uncertainty (jitter) has to be incorporated into the fitting function. A source for the jitter is the response of the SPAD. Ideally, the response of a perfect SPAD should be a delta function. However, the SPAD response curve exhibits a peak with a finite FWHM of 40\,ps. This response curve works as a point spread function $PSF(t)$ so that the measured histogram will be blurred. In addition, the temporal uncertainty of SAW-driven electrons, originating from uncertainty about the position of electrons in each SAW potential minimum, can also be another source of jitter. This jitter in the SAW-driven charge transport can be expressed as a Gaussian function
\begin{equation}
g(t,w)={\frac{1}{w \sqrt{2 \pi}}} \exp(- {\frac{1}{2}}{({\frac{t}{w}})}^2) \,,
\end{equation}
where $w$ is the parameter related to the FWHM according to FWHM $\sim 2.3 w$. Considering these sources of jitter, the time-resolved histogram $H(t)$ of the SAW-driven EL can be described as the convolution of the basic function $F$, the Gaussian function $g$, and the response curve $PSF$
\begin{equation}
H(t) = ((F \ast g) \ast PSF )(t) + \rmsub{BG}{EL} \,,
\label{eq_for_H}
\end{equation}
where $\rmsub{BG}{EL}$ is a constant for a constant background in the SAW-driven EL, which may result from long-lifetime excitons or after-pulsing of the SPADs. $H(t)$ is therefore determined by those previously mentioned parameters $t_0$, $\rmsub{t}{SAW}$, $\tau$, $w$, $A$ and $\rmsub{BG}{EL}$. Fig.\ \ref{fit_time_resolved_func} shows an example of $H(t)$. This function $H(t)$ is then used to fit the time-resolved EL data.

\subsection{S4. Fitting for Autocorrelation Histogram}

A function describing the autocorrelation histogram of a SAW-driven EL signal has to be established in order to fit the measured autocorrelation histogram. First, the autocorrelation $G(\Delta t)$ of a real signal $S(t)$ can be expressed as 
\begin{equation}
G(\Delta t) = \int S(u) S(u - \Delta t) du \,.
\end{equation}
This can be re-written as
\begin{align}
G(\Delta t) &= \int S(u) S(-(\Delta t - u)) du = \int S(u) \rmsub{S}{mirror}(\Delta t - u) du \notag \\ &= (S \ast \rmsub{S}{mirror})(\Delta t) \,,
\end{align}
where $\rmsub{S}{mirror}(t)$ is the mirror image of $S(t)$ so that $\rmsub{S}{mirror}(t) = S(-t)$. So, the autocorrelation $G(\Delta t)$ can also be understood as the convolution of $S(t)$ and its mirror image $\rmsub{S}{mirror}(t)$.

\begin{figure}
\centering    
	\includegraphics[width=0.7\textwidth]{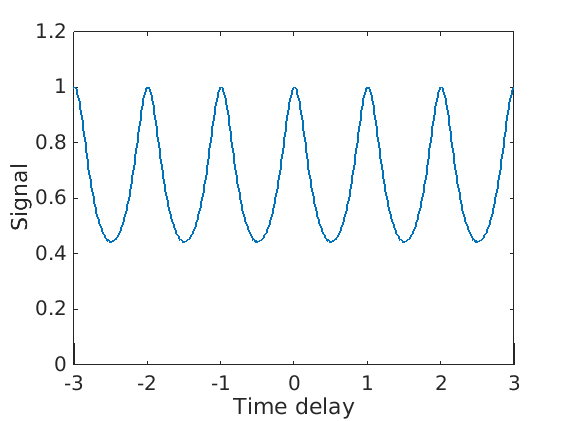}
\caption{Example of $G(\Delta t)$, which is the convolution of $H(t)$ in Fig.\ \ref{fit_time_resolved_func} and its mirror image $\rmsub{H}{mirror}(t)$.}
\label{multi_g2_function}
\end{figure}

In the case of an autocorrelation histogram of the SAW-driven EL, the function $G(\Delta t)$ becomes
\begin{equation}
G(\Delta t) = (H \ast \rmsub{H}{mirror})(\Delta t) \,.
\label{eq_G_function}
\end{equation}
where $H = ((F \ast g) \ast PSF ) + \rmsub{BG}{EL}$ (Equation~\ref{eq_for_H}). This means the autocorrelation histogram can be obtained by the convolution of the SAW-driven EL $H(t)$ and its mirror image $\rmsub{H}{mirror}(t)$. One such example is shown in Fig.\ \ref{multi_g2_function}. This function $G(\Delta t)$ can then be used to fit the measured autocorrelation histogram.

Because a constant background $\rmsub{BG}{EL}$ only causes a constant background $\rmsub{BG}{g2}$ in the autocorrelation, Equation~\ref{eq_G_function} can be re-written as
\begin{equation}
G(\Delta t) = (H' \ast \rmsub{H'}{mirror})(\Delta t) + \rmsub{BG}{g2} \,,
\label{eq_G_modified_1}
\end{equation}
where $H'(t) = ((F \ast g) \ast PSF )(t)$ is the function for the SAW-driven EL without $\rmsub{BG}{EL}$, and $\rmsub{H'}{mirror}(t)$ is its mirror image.

\subsection{S5. Extraction of the Second-order Correlation Function $g^{(2)}(\Delta t)$}

\begin{figure}
\centering    
\includegraphics[width=0.81\textwidth]{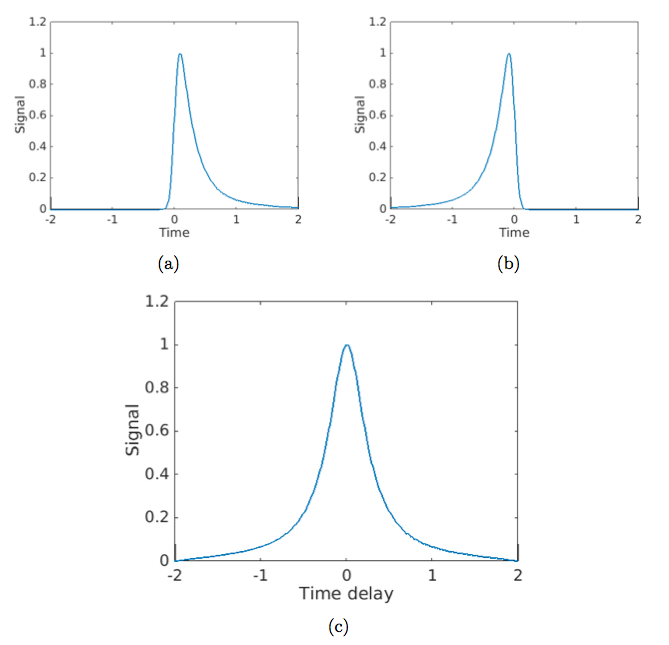}
\caption{(a) Example of $h(t)$ with $\tau = 0.2$, and $w = 0.05$. (b) Mirror image $\rmsub{h}{mirror}(t)$ of $h(t)$. (c) $\rmsub{g}{auto}(\Delta t)$, which is the convolution of $h(t)$ and $\rmsub{h}{mirror}(t)$.}
\label{single_g2_function}
\end{figure}

To calculate the contribution of one specific peak to the autocorrelation histogram, a theoretical function describing the shape of each peak also has to be established. Since $F$ is the summation of a series of equally-spaced function $f$ (see Equation~\ref{eq_F_function}), it can be shown that
\begin{align}
G(\Delta t) &= \sum \limits_{i = -n}^{n} (h \ast \rmsub{h}{mirror})(\Delta t + i \cdot \rmsub{t}{SAW}) + \rmsub{BG}{g2} \notag \\
&= \sum \limits_{i = -n}^{n} J(\Delta t + i \cdot \rmsub{t}{SAW}) + \rmsub{BG}{g2} \,,
\end{align}
where $n \rightarrow \infty$ is the number of EL peaks, and
\begin{equation}
h(t)=((f \ast g) \ast PSF )(t)\,.
\end{equation}
$h(t)$ can be understood as the EL signal from a SAW minimum. So, the autocorrelation histogram can be seen as the summation of a series of equally-spaced functions $J(\Delta t) = (h \ast \rmsub{h}{mirror})(\Delta t)$. Fig.\ \ref{single_g2_function}(a) shows an example of $h(t)$ with $\tau = 0.2$, and $w = 0.05$ while Fig.\ \ref{single_g2_function}(b) shows the mirror image $\rmsub{h}{mirror}(t)$. Their convolution $J(\Delta t)$ is plotted in Fig.\ \ref{single_g2_function}(c). This single peak can be understood as the actual shape of individual peaks in Fig.\ \ref{multi_g2_function}, which means that the contribution from a specific peak can be individually evaluated even though there is significant overlap between these peaks. So, if the theoretical function $J(\Delta t)$ for SAW-driven EL is known, the real signal from a suppressed peak, such as that in Fig.\ 3(a), can be estimated more accurately.

\begin{figure}
\centering    
\includegraphics[width=0.7\textwidth]{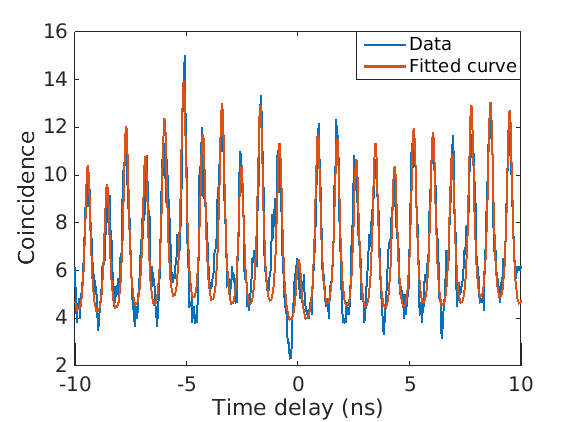}
\caption{Autocorrelation histogram in Fig.\ 3(a) and the best-fit curve using Equation~\ref{eq_g2_extract_function}.}
\label{g2_individual_fit}
\end{figure}

From the fit for the averaged histogram in Fig.\ 3(b), the shape of each peak is determined by $J(\Delta t)$ with $\tau = 99.6$\,ps, and $w = 33$\,ps, and $\rmsub{BG}{g2} = 2.79$. It can be assumed that each peak in Fig.\ 3(a) has the same shape but different peak amplitude due to the statistical sample variance. These peaks at $\Delta t= \rmsub{\Delta t}{(i)}$ have different amplitudes $\rmsub{A}{g2(i)}$, which are proportional to $g^{(2)}(\rmsub{\Delta t}{(i)})$. A modified autocorrelation function reflecting the variance can be expressed as
\begin{equation}
G'(\Delta t) = \sum \limits_{i = -\infty}^{\infty} \rmsub{A}{g2(i)} \cdot J(\Delta t + i \cdot \rmsub{t}{SAW}) + \rmsub{BG}{g2} \,.
\label{eq_g2_extract_function}
\end{equation}
In order to evaluate $g^{(2)}(\Delta t)$, $\rmsub{A}{g2(i)}$ at individual peaks have to be obtained by fitting the autocorrelation histogram to $G'(\Delta t)$ with $\tau = 99.6$\,ps, $w = 33$\,ps, and $\rmsub{BG}{g2} = 2.79$. Fig.\ \ref{g2_individual_fit} shows the data from Fig.\ 3(a), along with the best-fit curve for the suppressed peak and the nearest 22 peaks. 

The fitting process assigns to each peak an amplitude $\rmsub{A}{g2(i)}$. The second-order correlation function $g^{(2)}(\Delta t)$ is then obtained from the normalised $\rmsub{A}{g2(i)}$ as a function of time delay, as shown in Fig.\ 4(a).

\subsection{S6. Estimate of the Probability Distribution}

Because the number of electrons carried in a SAW minimum can be 0, 1, 2, 3 and so on, the probability $P_n$ of having $n$ electrons will follow a probability distribution. Since the average electron number $N_{\rm avg}$ in each SAW minimum is $0.89$, the probability of carrying more than four electrons in a SAW minimum should be negligible. Hence, the wave function $\lvert \psi \rangle$ for the electrons in a SAW minimum can be expressed as a superposition state
\begin{equation}
\lvert \psi \rangle =\sqrt{P_{0}}\lvert 0 \rangle+\sqrt{P_{1}}\lvert 1 \rangle+\sqrt{P_{2}}\lvert 2 \rangle+ \sqrt{P_{3}}\lvert 3 \rangle \,,
\end{equation}
with
\begin{equation}
\sum_{i} P_{i}=1 \  \text{and} \ 0 \leq P_{i} \leq 1 \,,
\label{eq_prob_sum}
\end{equation}
where $\lvert i \rangle$ is the $i$-electron state in the Fock basis, and $P_i$ is the probability of measuring $i$ electrons. The SAW-driven EL signal due to this electron-number probability distribution \{$P_i$\} then leads to the second-order correlation function $g^{(2)}(0)=0.39$. Hence, the probability distribution \{$P_i$\} in $\lvert \psi \rangle$ manifests $N_{\rm avg} = 0.89$ and $g^{(2)}(0) = 0.39$. $N_{\rm avg}$ and $g^{(2)}(0)$ can be expressed quantum mechanically as
\begin{equation}
N_{\rm avg}=\langle \psi \lvert \hat{a}^{\dagger} \hat{a} \rvert \psi \rangle=0.89 \,,
\label{eq_n_avg}
\end{equation}
and
\begin{equation}
g^{(2)}(0)=\frac{\langle \psi \lvert \hat{a}^{\dagger}\hat{a}^{\dagger}\hat{a} \hat{a}   \rvert \psi \rangle}{{\langle \psi \lvert \hat{a}^{\dagger} \hat{a} \rvert \psi \rangle}^{2}} = 0.39 \,,
\label{eq_g2_zero}
\end{equation}
where $\hat{a}$ and $\hat{a}^{\dagger}$ represent the annihilation and creation operators. Since $N_{\rm avg} = 0.89 < 1$, it should be safe to assume that a SAW minimum is less likely to contain two electrons than one electron, and is even more unlikely to contain three electrons. It seems likely, therefore, that
\begin{equation}
\frac{P_2}{P_1} \geq \frac{P_3}{P_2} \,.
\label{eq_p_ratio}
\end{equation}
If so, taking the equality $\frac{P_2}{P_1} = \frac{P_3}{P_2}$, and hence a larger value of $P_3$, and using Equations~\ref{eq_prob_sum} to~\ref{eq_g2_zero}, a probability distribution \{$P_0, P_1, P_2, P_3$\} $=$ \{$0.25, 0.63, 0.10, 0.02$\} can be obtained. This gives $P_1/(P_1 + P_{\rm multi})=P_1/(P_1 + P_2 + P_3)=0.84$. Alternatively, suppose that $P_3=P_2$, giving the highest possible value of $P_3$. Then $P_1=0.7$, which is even higher and gives $P_1/(P_1 + P_{\rm multi})=0.9$ (\{$P_0, P_1, P_2, P_3$\} $=$ \{$0.22, 0.7, 0.04, 0.04$\}). If, instead, $P_3=0$, then $P_1=0.58$, which gives $P_1/(P_1 + P_{\rm multi})=0.79$ (\{$P_0, P_1, P_2, P_3$\} $=$ \{$0.27, 0.58, 0.15, 0$\}). 

\subsection{S7. Modelling of SAW-driven Charge Transport}

\begin{figure}
\centering    
\includegraphics[width=0.7\textwidth]{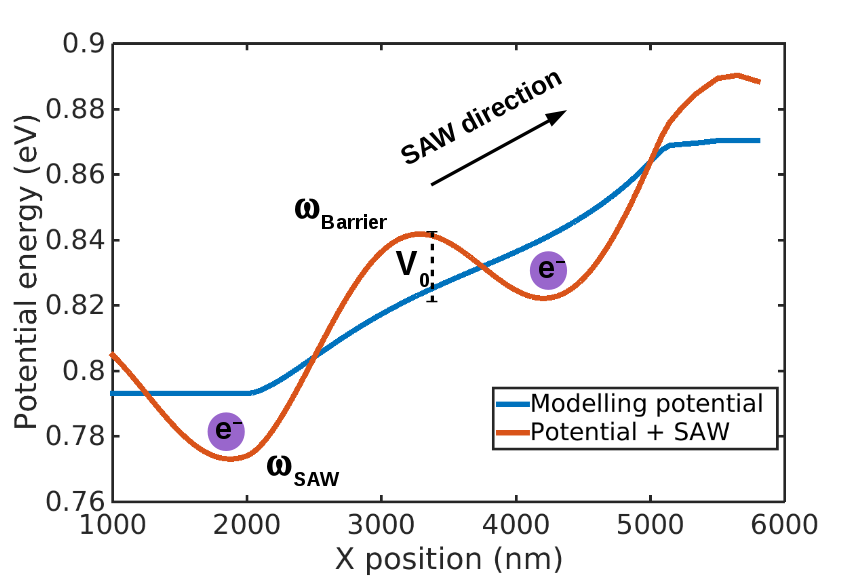}
\caption{Electrostatic potential of the conduction band (blue curve) of a SAW-driven \textit{n-i-p} junction, calculated using Nextnano. The red curve shows the electrostatic potential superposed by a SAW potential with SAW amplitude of 20\,meV. Electrons are trapped in the SAW minimums and dragged from the region of electrons ($x = 2000$\,nm) to the region of holes ($x = 5000$\,nm). Each SAW minimum works as a dynamic quantum dot (with potential height $V_0$, curvature at the top of the barrier $\rmsub{\omega}{Barrier}$, and curvature at the bottom of the quantum dot $\rmsub{\omega}{SAW}$).}
\label{fig_nextnano_saw}
\end{figure}

\begin{figure}[t!] 
\centering    
\includegraphics[width=0.7\textwidth]{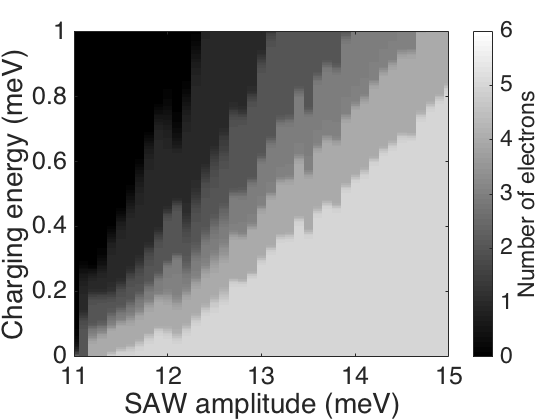}
\caption{Modelling result of the SAW-driven charge transport. Number of SAW-driven electrons arriving at the region of holes as a function of SAW amplitude and charging energy.}
\label{fig_tunnelling}
\end{figure}

In the main text, the non-zero probability of multiple occupation, $P_{\rm multi} \equiv P_2 + P_3 = 0.08$--$0.15$ when $N_{\rm avg}=0.89$, means that a SAW minimum may sometimes contain more than one electron and hence cause unwanted multi-photon emission. In order to understand qualitatively how to enhance the single-electron probability $P_1$, we have built a simplified model for the SAW-driven charge transport. To build the model, the first step is to calculate the electrostatic potential of the SAW-driven \textit{n-i-p} junction with S-D bias and gate voltages similar to the experiment. The blue curve in Fig.\ \ref{fig_nextnano_saw} shows the conduction band along the SAW-propagation direction (the $x$ axis), where the source (electron) bias $\rmsub{V}{source} = -0.8$\,V, the drain (hole) bias $\rmsub{V}{drain} = 0.65$\,V, and the side-gate voltage $\rmsub{V}{SiG} = -0.4$\,V. This calculation of the electrostatic potential, based on the real device geometry, was carried out using the partial differential equation solver Nextnano~\citep{Hou2018a,NextnanoGm}. As can be seen, the potential difference between the region of electrons (left) and the region of holes (right) is about 80\,meV, so electrons cannot overcome this potential difference and recombine with holes unless a SAW drags these electrons to the region of holes. The red curve in Fig.\ \ref{fig_nextnano_saw} shows the calculated conduction band with superposition of a SAW potential (amplitude $\rmsub{A}{SAW} = 20$\,meV). The SAW potential minima work as dynamic quantum dots, trapping the electrons and transporting them to the region of holes. During the transportation of electrons, some of them will tunnel out of the potential minima if the electrochemical potential $\rmsub{u}{(SAW dot)}$ of the dynamic quantum dots is close to or higher than the quantum-dot barrier height $V_0$ created by the SAW. The probability of tunnelling in one attempt can be described by the transmission probability of non-interacting electrons through a saddle-point potential~\citep{Buttiker1990,Astley2007}
\begin{equation}
T = \frac{1}{1+e^{-{\pi}{\epsilon}_n}}
\label{eq_tunnel_prob}
\end{equation}
with
\begin{equation}
{\epsilon}_n = 2[\rmsub{u}{(SAW dot)} - \hbar {\omega}_y(n_y+ \frac{1}{2}) - V_0)] / {\hbar}{\rmsub{\omega}{Barrier}} \,,
\end{equation}
where $\rmsub{\omega}{Barrier}$ is the curvature of the quantum-dot barrier expressed in frequency, ${\omega}_y$ is the curvature of the potential perpendicular to the SAW-propagation direction (the $y$ axis), which is related to the electrostatic potential of the etched 1D channel, and $n_y$ is the quantum number of the wave function on the $y$ axis. The 2D potential of a SAW-created quantum dot can be approximated as an anisotropic 2D harmonic oscillator with two different frequencies along the $x$ and $y$ axes
\begin{equation}
V(x,y) = {\frac{1}{2}}m{{{\omega}_x}^2}{x^2} + {\frac{1}{2}}m{{{\omega}_x}^2}{y^2}
\end{equation}
with energy states
\begin{equation}
E(n_x,n_y) = (\frac{1}{2} + n_y)\hbar{{\omega}_x} + (\frac{1}{2} + n_y)\hbar{{\omega}_y} \,,
\end{equation}
where ${\omega}_x = \rmsub{\omega}{SAW}$ (the curvature at the bottom of the SAW minimum) and $n_x$ is the quantum number of the wave function on the $x$ axis. Assuming (1) ${\omega}_x > {\omega}_y$ (which is true for the simulation result shown later) (2) the charging energy $\rmsub{E}{c}$ for adding one electron is dominated by the patterned gates rather than electron-electron interactions, and (3) spin-spin interactions can be ignored, then the electrochemical potential $\rmsub{u}{(SAW dot)}(N)$ for the first six electrons in the dot can be written as~\citep{Kouwenhoven2001}
\begin{align}
\rmsub{u}{(SAW dot)}(1) &= \rmsub{E}{c} + E(0,0) \notag \\
\rmsub{u}{(SAW dot)}(2) &= 2{\rmsub{E}{c}} + E(0,0) \notag \\
\rmsub{u}{(SAW dot)}(3) &= 3{\rmsub{E}{c}} + E(0,1) \notag \\
\rmsub{u}{(SAW dot)}(4) &= 4{\rmsub{E}{c}} + E(0,1) \notag \\
\rmsub{u}{(SAW dot)}(5) &= 5{\rmsub{E}{c}} + E(1,0) \notag \\
\rmsub{u}{(SAW dot)}(6) &= 6{\rmsub{E}{c}} + E(1,0) \,.
\label{eq_u_expression}
\end{align}
Therefore, the tunnelling probabilities of the first $N$ electrons in the SAW-created dot (with a given SAW amplitude and at a given dot position during transport) can be calculated using Equation~\ref{eq_tunnel_prob} once $\rmsub{\omega}{Barrier}$, $\rmsub{\omega}{SAW}$, $\omega_y$, and $V_0$ are obtained from fitting the conduction band with a SAW superposed at the corresponding phase. In Fig.\ \ref{fig_tunnelling}, in order to observe the overall trend in the modelling result, a few poorly-fitted $V_0$ and $\rmsub{\omega}{Barrier}$ are removed and interpolated. Subsequently, a probability distribution of electron numbers in the dot can be calculated by the evolution of electron population, due to tunnelling, from an initial state (e.g. 5 electrons at the beginning) in the region of electrons ($x = 2000$\,nm in Fig.\ \ref{fig_nextnano_saw}) to the final state in the region of holes ($x = 5000$\,nm).

The modelling result in Fig.\ \ref{fig_tunnelling} shows the average number of electrons corresponding to the final probability distribution at $x = 5000$\,nm as a function of SAW amplitude $\rmsub{A}{SAW}$ and charging energy $\rmsub{E}{c}$, where the initial state is 5 electrons at $x = 2000$\,nm. As can be seen, at a fixed $\rmsub{E}{c}$, when $\rmsub{A}{SAW}$ decreases, the number of electrons also decreases from 5 to 0 due to weakening of the confinement around the SAW minimum. In addition, at a higher $\rmsub{E}{c}$, the steps between different electron numbers are wider in terms of $\rmsub{A}{SAW}$ and are more well-defined (with a higher contrast). This indicates that a more well-defined electron number, and thus a purer single-photon emission, may be achieved by increasing the charging energy. This requires a stronger confinement to be provided laterally by the etching and perhaps also longitudinally by the SAW potential.

\subsection{S8. Future Improvements}

Regarding the EL efficiency $\eta$, as electrons are carried in the SAW minima, they may pass the \textit{p} region without immediately recombining with holes near the junction, and eventually produce photons far away from the junction, which cannot be collected by the confocal optics. The existence of such electrons may be one of the reasons for the low $\eta=2.5\%$. Hence, capturing electrons using etching or extra electrostatic potentials around the junction may be helpful for improving $\eta$. Having a Bragg stack beneath the quantum well and building a pillar with an optical cavity designed to be resonant with the neutral-exciton energy will also help the photons to be emitted into a desired mode, and thus increase $\eta$. Since the neutral-exciton energy in the quantum well should be reproducible between wafers, and can be characterised using photoluminescence (PL), the Bragg stack and the optical cavity can be deterministically designed for the expected photon energy, which makes the brightness enhancement more reliable than for self-assembled InGaAs QDs.

Regarding photon indistinguishability, the total spectral FWHM of the EL signal is caused by homogeneous (natural linewidth and phonon scattering) and inhomogeneous (local electric field and interface roughness) broadening. It was found that, in a high-quality 15\,nm GaAs quantum well, the homogeneous broadening is about 100-150\,$\mu$eV (FWHM) (with an acoustic-phonon broadening coefficient $\gamma \simeq 3$\,$\mu$eV/K), and the inhomogeneous broadening is about 100\,$\mu$eV, at 8\,K~\citep{Srinivas1992}. Therefore, if a device can be made in a high-quality quantum well with improved interface roughness, the spectral FWHM from the SAW-driven single-photon source may be reduced to about 80\,$\mu$eV at 1.5\,K, which will improve the photon indistinguishability. For applications like quantum repeaters and photonic quantum computation, two photons sent to a beam-splitter are required to remain in phase for the duration of the overlapping photon wavepackets in order to maximise photon-photon interference. For two photons with an energy difference $\Delta E_{\rm ph}$ and a wavepacket size $\Delta t_{\rm ph}$ ($\sim$ carrier lifetime $\tau$) in the time domain, the requirement can be expressed as $\frac{\Delta E_{\rm ph}}{\hbar} \Delta t_{\rm ph} \leq \pi$. If $\Delta E_{\rm ph} = 80$\,$\mu$eV, then $\Delta t_{\rm ph}$ needs to be shorter than 25\,ps, which may be achievable by increasing the recombination rate or reducing the SAW-transport jitter.

\bibliography{supplementary_inf.bbl}